\documentclass[aip,jcp,amsmath,amssymb,reprint,superscriptaddress]{revtex4-2}
\usepackage{graphicx}
\usepackage{epsf}
\usepackage{bm}
\usepackage{physics}
\usepackage{lipsum}
\usepackage[normalem]{ulem}
\usepackage{txfonts}
\usepackage{booktabs}
\usepackage{multirow}
\usepackage{xr}

\graphicspath{{./}{./figs/}{../figs/}}
\begin{document}

\title{
Deep learning of committor for ion dissociation and interpretable
analysis of solvent effects using atom-centered symmetry functions
}

\author{Kenji Okada}
\affiliation{Division of Chemical Engineering, Department of Materials Engineering Science, Graduate School of Engineering Science, The University of Osaka, Toyonaka, Osaka 560-8531, Japan}

\author{Kazushi Okada}
\affiliation{Division of Chemical Engineering, Department of Materials Engineering Science, Graduate School of Engineering Science, The University of Osaka, Toyonaka, Osaka 560-8531, Japan}

\author{Kei-ichi Okazaki}
\email{keokazaki@ims.ac.jp}
\affiliation{Research Center for Computational Science, Institute for Molecular Science, Okazaki, Aichi 444-8585, Japan}
\affiliation{Graduate Institute for Advanced Studies, SOKENDAI, Okazaki,
Aichi 444-8585, Japan}

\author{Toshifumi Mori}
\email{toshi\_mori@cm.kyushu-u.ac.jp}
\affiliation{Institute for Materials Chemistry and
Engineering, Kyushu University, Kasuga, Fukuoka 816-8580, Japan}
\affiliation{Interdisciplinary Graduate School of Engineering Sciences,
Kyushu University, Kasuga, Fukuoka 816-8580, Japan}

\author{Kang Kim}
\email{kk@cheng.es.osaka-u.ac.jp}
\affiliation{Division of Chemical Engineering, Department of Materials Engineering Science, Graduate School of Engineering Science, The University of Osaka, Toyonaka, Osaka 560-8531, Japan}

\author{Nobuyuki Matubayasi}
\email{nobuyuki@cheng.es.osaka-u.ac.jp}
\affiliation{Division of Chemical Engineering, Department of Materials Engineering Science, Graduate School of Engineering Science, The University of Osaka, Toyonaka, Osaka 560-8531, Japan}

\date{\today}

\begin{abstract}
The association and dissociation of ion pairs in water are fundamental
 to 
 physical chemistry, yet their reaction coordinates are 
 complex, involving not only interionic distance but also
 solvent-mediated hydration structures. 
These processes are often
 represented by free-energy landscapes constructed from collective
 variables (CVs), such as 
 interionic distance and water bridging structures; however, it remains
 uncertain whether such representations reliably capture the transition
 pathways between the two associated and dissociated states.
In this study, we employ deep learning to identify reaction
 coordinates for NaCl ion pair association and dissociation in water,
 using the committor as a quantitative measure of progress along the transition
 pathway through the transition state.
The solvent environment surrounding the ions is encoded
 through descriptors based on atom-centered symmetry functions (ACSFs),
 which serve as input variables for the neural network.
In
 addition, 
Shapley Additive exPlanations analysis, as 
an explainable artificial intelligence technique, is applied to identify 
ACSFs that contribute to the reaction coordinate.
A comparative analysis of their correlation with CVs representing water
 bridging structures, such as interionic water density and the number of
 water molecules coordinating both ions, 
further provides a
molecular-level interpretation of the ion association-dissociation mechanism in
 water.
\end{abstract}
\maketitle

\section{Introduction}

To understand transition processes involving many degrees of freedom in
complex molecular systems, such
as ion dissociation and isomerization, it is essential to
examine the corresponding
free-energy landscape.~\cite{chipot2007Free, zuckerman2010Statistical,
peters2017Reaction, pietrucci2017Strategies}
From the system's total degrees of freedom,
a collective variable (CV) $X$, 
for example, an interatomic distance, bond angle, or dihedral angle, is
conventionally employed.
The free-energy landscape can be described by the potential of mean
force (PMF), $-k_\mathrm{B}T\ln P(X)$, 
where $P(X)$
denotes the probability distribution function with respect to $X$.
When the resulting PMF exhibits two minima
separated by a saddle point, these minima correspond to distinct stable
states, while the transition pathway is expected to pass in the vicinity
of the saddle point. 
In this context, the saddle point represents the transition state
(TS), and the CV $X$
can be regarded as the reaction coordinate (RC). 
However, a fundamental difficulty arises in that the selection of 
$X$ is arbitrary.

A method known as committor analysis is commonly used to evaluate CV as
an appropriate RC.~\cite{pande1998Pathways, bolhuis2000Reaction, bolhuis2002TRANSITION,
hummer2004Transition, 
best2005Reaction, e2005Transition, jung2017Transition, rogal2021Reaction, bolhuis2021Transition} 
The committor, $p_\mathrm{B}^*$, 
is defined as the probability that a trajectory initiated from a given
configuration reaches the product state B
before reaching the reactant state A.
A value of $p_\mathrm{B}^*\approx 0$
indicates that the configuration lies near state A, whereas
$p_\mathrm{B}^*\approx 1$ signifies proximity to state B.
Configurations for which $p_\mathrm{B}^*=0.5$ 
correspond to the TS ensemble. 
When a CV is chosen such that configurations with $p_\mathrm{B}^*=0.5$ 
coincide with the saddle point on the PMF, that CV can
be regarded as the appropriate RC. 
However, because a large number of possible CVs exist and the evaluation
is typically performed by trial and error, an automated and systematic
approach is required.~\cite{li2014Recent}

Machine learning approaches based on committor analysis have shown
considerable promise for identifying RC, leading to
the development of various methodological
frameworks.~\cite{ma2005Automatic, peters2006Obtaining,
peters2007Extensions, peters2016Reaction, schneider2017Stochastic,
rogal2019NeuralNetworkBased, mori2020Learning, bonati2020DataDriven, 
bonati2021Deep, frassek2021Extended, magrino2022Critical,
neumann2022Artificial, lazzeri2023Molecular,
jung2023Machineguided, chen2023Discovering, naleem2023Exploration, france-lanord2024DataDriven,
zhang2024DescriptorFree,
herringer2024Permutationally, zhu2025Enhanceda, megias2025Iterative, dietrich2025Reproducibility}
Recently, we developed a deep learning framework trained on committor
values $p_\mathrm{B}^*$ of configurations sampled near the saddle point
region from molecular dynamics (MD) simulations.~\cite{kikutsuji2022Explaining}
In this approach, candidate CVs are employed as input features, and the
corresponding RC is predicted as the output variable through the neural
network using $p_B^*$ as the learning target.
Because deep learning models often behave as black boxes, eXplainable AI
(XAI) techniques are further employed to quantify the contribution of
each input variable to the predicted RC. 
This approach enables the identification of CVs with significant
contribution, providing a representation of the TS on the PMF
that is consistent with the underlying transition pathways.

As an application, we previously investigated the isomerization reaction of alanine
dipeptide and identified dihedral angles that primary 
contributor to the RC.~\cite{kikutsuji2022Explaining}
Furthermore, we demonstrated that the isomerization can be characterized
not only by the dihedral angles but also by specific interatomic
distances.~\cite{okada2024Unveiling}
We also investigated how hyperparameter tuning in the neural network
model influences the performance and reliability of the
identified RC.~\cite{kawashima2025Investigating}

In this study, we applied an explainable deep learning-based analytical framework for
identifying appropriate CVs serving as RC to the
association and dissociation process of NaCl ion pair in water.
This process may appear to be described adequately by the interionic distance 
$r_\mathrm{ion}$.~\cite{karim1986Dynamics, karim1986Ratea,
ciccotti1989Constrained, ciccotti1990Dynamics, guardia1991Potentiala,
rey1992Dynamical, smith1994Computer, brunig2022PairReaction}
Geissler \textit{et al.} applied committor analysis to the
dissociation of a NaCl ion pair
in water.~\cite{geissler1999Kinetic}
They constructed 
a TS ensemble by constraining $r_\mathrm{ion}$
to the saddle point region of the PMF.
The resulting distribution of the committer $p_\mathrm{B}^*$ was
bimodal, with pronounced peaks at 
$p_\mathrm{B}^*=0$ and $p_\mathrm{B}^*=1$, while the
distribution near $p_\mathrm{B}^*=0.5$ was relatively low.
This result indicated $r_\mathrm{ion}$ alone does not fully capture the
TS along the RC.
They further examined the structural features of the solvent environment
surrounding the ions for configurations with $p_\mathrm{B}^*=0$ and $p_\mathrm{B}^*=1$.
The analysis showed that Na ions were predominantly five coordinated at $p_\mathrm{B}^*=0$, 
whereas six coordination dominated at $p_\mathrm{B}^*=1$.
In contrast, no substantial change was observed in the coordination
environment of Cl ions.
These findings suggest that the dissociation process involves an increase in the
coordination number of Na ions from the associated to the dissociated
state.

Furthermore, Bellard and Dellago performed committor analysis, 
in which both $r_\mathrm{ion}$ and water molecules were
constrained.~\cite{ballard2012Mechanism}
In their approach, water molecules were restricted by varying the probe
region surrounding the ions, and the resulting committor distributions
were examined. 
The results showed that, as the constrained region was extended from the
first to the third coordination shell, the committor distribution
progressively sharpened around $p_\mathrm{B}^*=0.5$.
This result demonstrated that ion pair dissociation is closely
associated with solvent motion between the second and third solvation shells.

Here, particular attention is devoted to the construction of
CVs that describe water molecules surrounding ions.~\cite{mullen2014Transmission,
yonetani2015Distinct,yonetani2017Solventcoordinate, salanne2017Ca2+Cl,
joswiak2018Ion, oh2019Understanding, zhang2020Dissociation,
wang2022Influence, jung2023Machineguided, wilke2025NaCl} 
M\"{u}llen \textit{et al.} calculated a total of 71 CVs, including 
the local density of
solvent molecules or atoms relative to the ion pair such as 
first and second shell coordination numbers, interionic water density,
energy gaps, and other electrostatic
descriptors.~\cite{mullen2014Transmission}
Then, the combinations of up to three CVs were
employed in the Inertial Likelihood Maximization method.~\cite{peters2012Inertial}
In particular, CVs such as the interionic water density
and the water molecules simultaneously coordinating both ions were identified as
important.

Jung \textit{et al.}
developed a committor based machine learning approach combined with 
symbolic regression and applied it to various systems, including ion
dissociation in water.~\cite{jung2023Machineguided}
They employed atom-centered symmetry functions
(ACSFs), widely used as descriptors in machine learning
potentials,~\cite{behler2007Generalized, behler2011Atomcentered, behler2016Perspective} as
CVs.
Specifically, they incorporated ACSFs to describe the geometric
structures of water O and H atoms surrounding cations and anions.
Symbolic regression enables the RC identified by neural network learning
of the committor $p_\mathrm{B}^*$
to be expressed as explicit functional forms of the primary contributing CVs, including
the ACSFs that describe the structure of water O atoms around the cations.

In this study, we aim to enhance
interpretability of ACSF-based committor models trained by a neural network.
In particular, we employ SHapley Additive exPlanations (SHAP), a
game-theory-based method that decomposes a trained model prediction into
additive contributions from individual input features.~\cite{lundberg2017unified}
Here, SHAP is used to assess the relative importance of candidate
CVs and to elucidate how they contribute to the committor
prediction learned by the neural network. 
Our primary objective is not
the derivation of an explicit functional form relating the input
variables to the prediction, but rather the physical interpretation
of the ACSFs highlighted by the SHAP analysis.
In this sense,
our approach complements existing efforts based on symbolic regression,
going beyond the derivation of compact analytical expressions by
focusing on the physical meaning and mutual consistency of the
contributing ACSFs. 
To this end, we perform a comparative analysis
between the ACSFs emphasized by SHAP and conventional CVs, such as the
interionic water density
and the number of water molecules that simultaneously coordinate both
ions, introduced in previous studies. 
This approach allows us to
examine consistency with existing physical understanding, clarify
differences, and extract additional insights afforded by the data-driven
interpretation.

\section{Methods}

\subsection{MD simulations and committor evaluation}

\begin{figure}[t]
    \centering
    \includegraphics[width=0.8\linewidth]{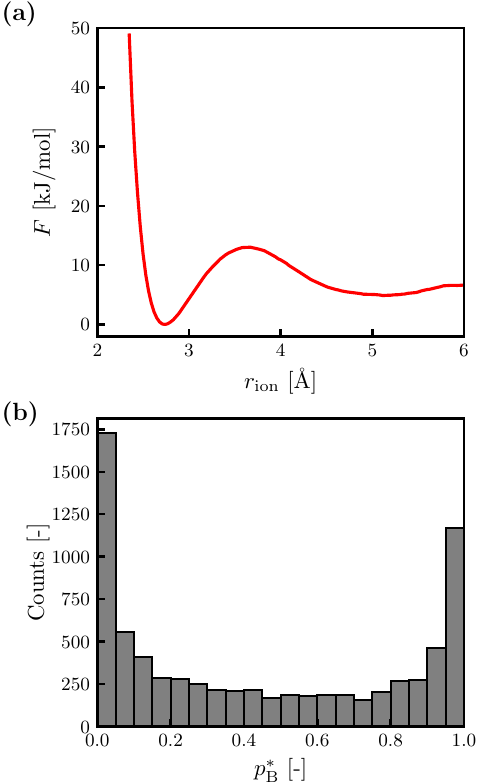}
    \caption{(a) PMF $F(r_\mathrm{ion})$ as a function the interionic distance $r_\mathrm{ion}$.
The point with the minimum energy is set to 0 kJ/mol.
(b) Distribution of Committor $p_\mathrm{B}^*$ evaluated for
 configurations sampled within the range $3.2~{\AA} < r_\mathrm{ion} < 4.3$ {\AA}.
}
     \label{fig:pmf}
\end{figure}

MD simulations were performed for NaCl ion pair in
water.
The system had a linear dimension of 40 {\AA} and was simulated under
periodic boundary conditions.
It comprised one pair of Na and Cl ions by the AMBER99 force
field~\cite{wang2000How} and 2048 TIP/4P-Ew~\cite{horn2004Development}
water molecules. 
Initially, MD simulations using the umbrella sampling method were
performed at 300 K
for 10 ns with the time step of 2 fs.
A harmonic bias with a spring constant of $k=500$ kJ mol$^{-1}$
{\AA}$^{-2}$ was applied at 30 window positions spaced at 0.2 {\AA} intervals
over the interion distance range from 2.2 {\AA} to 8.0 {\AA}.
The temperature was controlled using the Nos\'{e}--Hoover thermostat.
The real space cut-off length was set to 12 {\AA}, and long range electrostatic
interactions were evaluated using the particle mesh Ewald method.
The potential of mean force (PMF) as a function of $r_\mathrm{ion}$ was
then constructed using the WHAM method.
The resulting PMF is illustrated in Fig.~\ref{fig:pmf}(a).

Next, an additional umbrella sampling was performed centered at 
$r_\mathrm{ion}=3.65$ {\AA} 
with a harmonic bias of a spring constant $k=100$ kJ mol$^{-1}$
{\AA}$^{-2}$.
A total of 7600 configurations were extracted for 
interionic distances between 
$3.2~{\AA} < r_\mathrm{ion} < 4.3$ {\AA}.
This distance range was chosen to cover the saddle point position at 
$r_\mathrm{ion} =3.6$ {\AA} on the PMF,
as shown in Fig.~\ref{fig:pmf}(a).
The associated state A
is defined for $r< 3.2~{\AA}$, while the dissociated state 
B is defined for $r> 4.3~{\AA}$.
The committor $p_\mathrm{B}^*$
for each sampled configuration was evaluated by 
performing 100 independent 1 ps MD simulations at 300 K, each
initialized with velocities randomly 
drawn from the Maxwell--Boltzmann distribution for the sampled configuration.
The distribution of the committor $p_\mathrm{B}^*$ is shown in Fig.~\ref{fig:pmf}(b).
It can be seen that 
$p_\mathrm{B}^*$ tends toward 0 and 1, without exhibiting a pronounced
peak near 0.5. 
This result is consistent with the findings of previous
work,~\cite{geissler1999Kinetic} 
indicating that 
the CV of $r_\mathrm{ion}$ 
is insufficient to account for the TS.
All the MD simulations were performed using
GROMACS 2024.5.~\cite{abraham2015GROMACS}

\subsection{Neural network learning and SHAP}

The neural network used in this study consists of five hidden layers,
with the odd-numbered layers containing 400 nodes and even-numbered layers
containing 200 nodes. 
The leaky rectified linear unit (Leaky ReLU) was used as the activation
function, with the leaky parameter set to 0.01.
Candidate CVs are provided as the input features, and the output is a
one-dimensional variable $q$.
The network is trained to regress the relationship between 
$q$ and 
the committor $p_\mathrm{B}^*$ onto the sigmoid function
$p_\mathrm{B}(q)=(1+\tanh(q))/2$.
Consequently, $q$ serves as the RC.
The architecture of the neural network is same as that used in our
previous studies.~\cite{kikutsuji2022Explaining, okada2024Unveiling}

The cross-entropy between $p_\mathrm{B}(q)=(1+\tanh(q))$ and
$p_\mathrm{B}^*$ was employed as the loss function, which is expressed
as
\begin{align}
\mathcal{H} (p_\mathrm{B}, p_\mathrm{B}^*) &= - \sum_{k} p_\mathrm{B}^* (\bm{r}_k) \ln p_\mathrm{B}(q)\nonumber\\
&\qquad - \sum_{k} (1-p_\mathrm{B}^*(\bm{r}_k)) \ln [1- p_\mathrm{B}(q)],
\label{eq:cross_entropy}
\end{align}
where $\bm{r}_k$ denotes the $k$-th initial
configuration.~\cite{mori2020Dissecting, mori2020Learning}
This equation is derived from the Kullback--Leibler divergence, which
quantifies the discrepancy between the distribution of the committor
$p_\mathrm{B}^*$ and the expected committor function $p_\mathrm{B}(q)$.~\cite{mori2020Dissecting}
It is also be noted that Eq.~\eqref{eq:cross_entropy} represents a
generalization of the log-likelihood function.~\cite{peters2007Extensions}
The dataset of candidate CVs and committor $p_\mathrm{B}^*$
from 7600 coonfigurations was
partitioned into training, validation, and test datasets at a ratio of 5:1:4.
Furthermore, overfitting was prevented using Dropout (probability 
0.5) and L2 regularization (coefficient 0.001). 
Training was performed using the AdaMax optimizer with a batch size of
256 and a learning rate of  1 $\times 10^{-5}$, over 4000 epochs.
The epoch dependent loss functions of training and test datasets are
shown in Fig.~S1 of the Supplementary Material.

To enhance the interpretability of the neural network model, we employed
the SHAP
analysis as an XAI technique.~\cite{lundberg2017unified}
As described in the Introduction, SHAP quantifies the contribution of each input feature to the
predictions, thereby providing an interpretable decomposition of feature contribution.
In particular, SHAP employs an additive feature attribution framework using 
Shapley values based on game theory, ensuring a fair and
consistent distribution of prediction values among input features.
In the present study, this analysis is used to assess 
the relative importance of 
CVs for the committor prediction learned by the
neural network.
Note that while SHAP decomposes model predictions into
additive feature contributions, it does not provide an explicit
functional form of the relationship between the input variables
and the prediction, which is instead addressed separately by symbolic regression.~\cite{jung2023Machineguided}

\subsection{Input features as candidate CVs}

The ACSF describes how many atoms are located at specific distances and
angles around a central atom.~\cite{behler2011Atomcentered} 
Owing to their invariance under translational and
rotational operations, ACSFs provide a systematic and comprehensive 
representation of the solvent environment.
Note that Geiger and Dellago proposed a machine learning method using
ACSFs to classify the local structures of amorphous and crystalline
structures.~\cite{geiger2013Neural}
Extended methods for identifying local structures in MD simulations,
based on the neural network framework developed by Geiger and Dellago,
have also been proposed.~\cite{defever2019Generalized, fulford2019DeepIce}

To represent the solvent environment around the ions, 
we employed two types of ACSFs, $G_i^2$ and $G_i^5$ for
reference atom $i$, as input variables for neural networks.
These functions are defined as
\begin{align}
G_i^{2, Z_1} &= \sum_{j\ne i}^{|Z_1|} e^{-\eta(R_{ij}-R_\mathrm{s})^2} \cdot
 f_\mathrm{c}(R_{ij}), \\
G_i^{5, Z_1, Z_2} &= 2^{1-\zeta} \sum_{j\ne i}^{|Z_1|} \sum_{k\ne
 i}^{|Z_2|} (1+\lambda \cos \theta_{ijk})^\zeta \nonumber\\
&\qquad \cdot e^{-\eta((R_{ij}-R_\mathrm{s})^2+(R_{ik}-R_\mathrm{s})^2)} 
 \cdot f_\mathrm{c}(R_{ij}) \cdot
 f_\mathrm{c}(R_{ik}), 
\end{align}
respectively.
Note that $R_{ij}$ ($R_{ik}$) represents the distance between atoms $i$
and $j$ ($k$), and 
$\theta_{ijk}$ denotes the angle formed by atoms $j$ and $k$ with 
atom $i$ at the center.
$Z_1$ and $Z_2$ represent atomic species
surrounding the reference atom 
and the absolute value notation denotes their
respective atomic numbers.
Furthermore, the cut-off function is introduced as
\begin{equation}
f_\mathrm{c} (R_{ij}) = 
\begin{cases}
0.5\cdot \left[\cos(\pi R_{ij}/R_\mathrm{c})+1\right]  
& \text{for $R_{ij} < R_\mathrm{c}$} \\
0 
& \text{for $R_{ij} > R_\mathrm{c}$},
\end{cases}
\end{equation}
with the cut-off radius, $R_\mathrm{c}$.

The ACSF $G_i^2$
is defined as the sum of a Gaussian function weighted by the cutoff
function. 
The width of the Gaussian is determined by the parameter $\eta$, 
while its center can be positioned at a specific radial distance using
the parameter $R_\mathrm{s}$.
This function is particularly suitable for characterizing the spherical
shell environment surrounding a reference atom.
The ACSF $G_i^5$ extends $G_i^2$ by incorporating an additional target
atom and including an angular component. 
The parameter $\lambda$ can take values of $1$ or $-1$, shifting the
maximum of the cosine function to $\theta_{ijk}=0^\circ$ and
$\theta_{ijk}=180^\circ$, respectively.
The parameter $\zeta$
controls the angular resolution, with larger values of $\zeta$
corresponding to a narrower range of nonzero contributions in the ACSF.

As outlined in the Introduction, Jung \textit{et al.}
introduced 
ACSFs to describe the geometric structures of water
O and H atoms around cations anions 
as candidate CVs in their machine learning analysis of the committor.~\cite{jung2023Machineguided}
To consider CVs more comprehensively, 
we calculated all possible atomic
combinations as follows: For $G^2$, 
four combinations were used: ($i$-$Z_1$)=
(Na-O), (Na-H), (Cl-O), (Cl-H). 
For $G^5$, 
ten combinations were considered: 
($Z_1$-$i$-$Z_2$)=(O-Na-O), (O-Na-H), (H-Na-H), (Cl-Na-H),
(Cl-Na-O), (O-Cl-O), (O-Cl-H), (H-Cl-H), (Na-Cl-H), (Na-Cl-O).
By systematically varying other parameters, $R_\mathrm{s}$ (ranging from
1 {\AA} to 9 {\AA}), $\lambda$ (1 and -1),
and $\zeta$ (1, 2, 4, 8, 16, 32, and 64), we generated a total of 1296
descriptors, comprising 36 $G^2$ and 1260 $G^5$.
The cutoff radius was fixed at 
$R_\mathrm{c}=10.0$ {\AA}, with $\eta=2.0$ {\AA$^{-2}$} for
$G^2$, and $\eta=1.2$ {\AA$^{-2}$} for $G^5$.
Detailed definitions of the input features
of two ACSFs,
$G^2$ and $G^5$, are provided in Tables S1 of the Supplementary
Material.
All input variables for the neural network
were standardized.

\begin{figure}[t]
    \centering
    \includegraphics[width=0.8\linewidth]{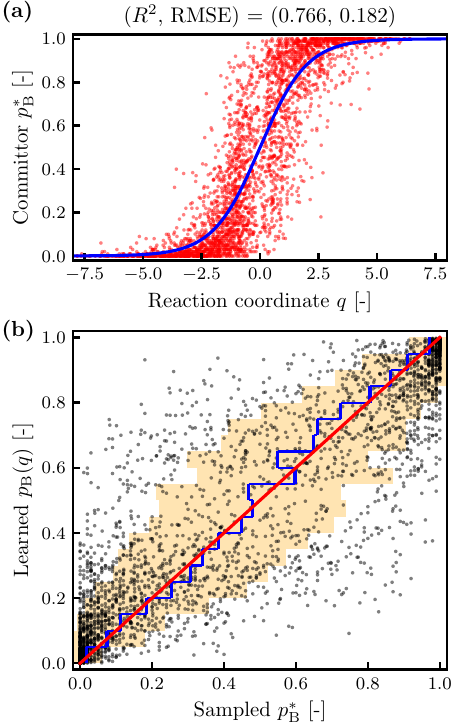}
    \caption{(a) Relationship between committor $p_\mathrm{B}^*$ and the
 RC $q$ predicted by the neural network trained model using the test dataset (3040 points).
The blue curve represents the sigmoidal function, $p_\mathrm{B}(q)=(1+\tanh(q))/2$.
(b) Comparison between the sampled committor $p_\mathrm{B}^*$ using the test dataset (3040 points)
 and the learned committor by the neural network.
The vertical axis is divided into 20 bins; the mean of each bin
 shown as a blue line. 
The orange shaded area indicates the standard deviation, and the red line represents the identity.
}
     \label{fig:committor}
\end{figure}

\section{Results and discussion}

\subsection{Predicted RC}

We first present the result of the neural network training on the committor $p_\mathrm{B}^*$.
Figure~\ref{fig:committor}(a) shows the relationship between 
$p_\mathrm{B}^*$ and $q$
using the test data. 
The model performance was evaluated using two metrics: the coefficient
of determination $R^2$ and 
the root-mean-square error (RMSE), which were found to be $(R^2, \text{RMSE}) =(0.766,0.128)$.
Our previous study investigated the isomerization of alanine dipeptides
using the same neural network architecture;
however, the present model exhibited slightly reduced predictive
performance relative to the earlier one.~\cite{kikutsuji2022Explaining,
okada2024Unveiling} 
This decrease is likely attributable to the substantial increase in
degrees of freedom arising from the inclusion of solvent, although we
consider the training to have been sufficiently converged.

Figure~\ref{fig:committor}(b) compares 
the learned committor by the neural network 
with the sampled committor $p_\mathrm{B}^*$.
The close quantitative agreement between the two indicates the neural
network accurately characterizes the reaction coordinate $q$, consistent
with previous observations by 
Jung \textit{et al}.~\cite{jung2023Machineguided}
In addition, Fig.~S2 of the supplementary material shows the
distribution of $p_\mathrm{B}^*$ near the TS at $q = 0$. 
Using $\alpha$ as a threshold, the distribution of $p_\mathrm{B}^*$
within the range $|q| < \alpha$ increasingly exhibits a tendency to peak
around 0.5 as $\alpha$ is reduced, although the peak is less sharply
defined than that observed
for the isomerization of
alanine dipeptides.~\cite{kikutsuji2022Explaining}
Nevertheless, the predicted 
$q$
obtained from the neural network model can serve as an appropriate RC
for ion pair dissociation.

\begin{figure*}[t]
    \centering
    \includegraphics[width=0.7\linewidth]{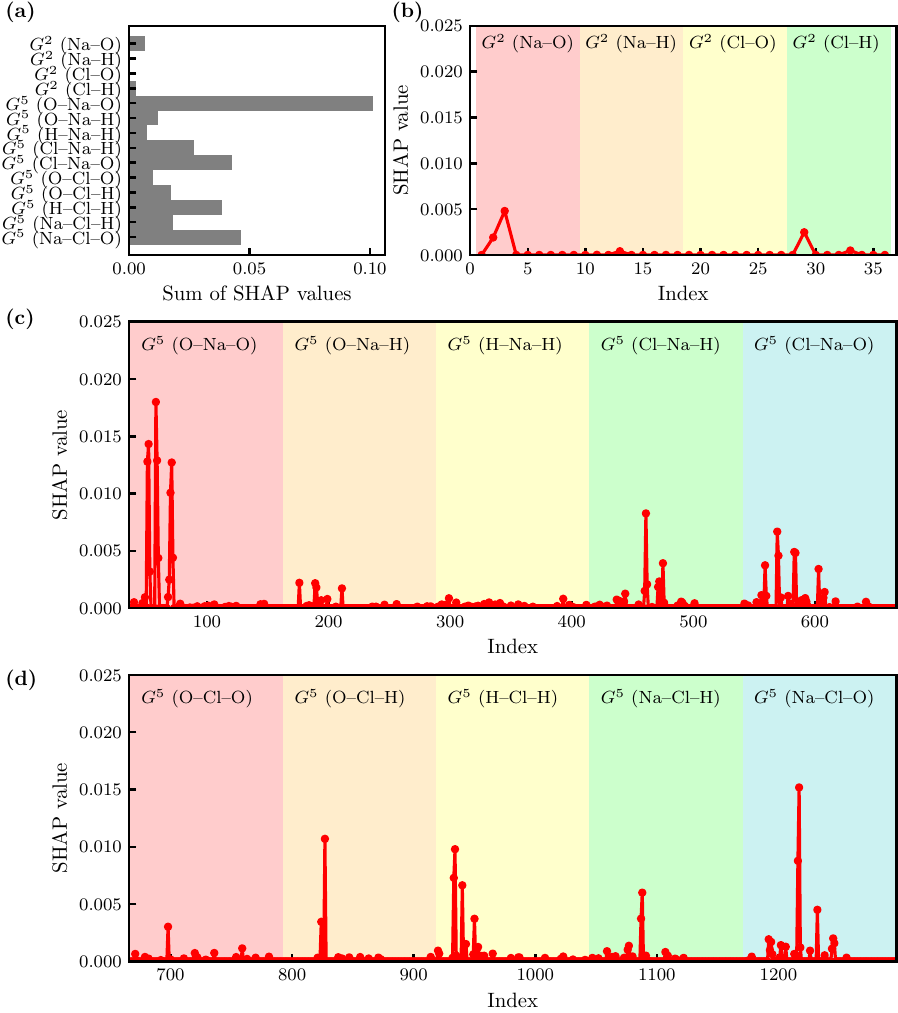}
    \caption{Feature contribution of each CV evaluated by the absolute SHAP value.
(a): Summed SHAP values for each atom combination ($i$-$Z_1$) in
$G^2$ and ($Z_1$-$i$-$Z_2$) in $G^5$ descriptors.
(b): Index dependence of SHAP values for $G^2$ descriptors.
(c) and (d): Index dependence of SHAP values for $G^5$ descriptors.
}
     \label{fig:SHAP}
\end{figure*}

\begin{table}[t]
\caption{Top five dominant indices, those parameters, and absolute SHAP
 values among $G^5$ (O-Na-O) descriptors.
}
\centering
\begin{tabular}{c c c c c}
\toprule
\midrule
index &  $R_\mathrm{s}$ [{\AA}] &  $\lambda$ & $\zeta$ & absolute SHAP value\\
\midrule
58 & 2.0 & -1 & 1 & 0.0180 \\
52 & 2.0 & 1 & 2 & 0.0143 \\
59 & 2.0 & -1 & 2 & 0.0129 \\
51 & 2.0 & 1 & 1 & 0.0128 \\
71 & 3.0 & 1 & 32 & 0.0127 \\
\midrule
\bottomrule
\end{tabular}
\label{table:SHAP_O-Na-O}
\end{table}

\begin{table}[t]
\caption{Top five dominant indices, those parameters, and absolute SHAP
 values among $G^5$ (Na-Cl-O) descriptors.
}
\centering
\begin{tabular}{c c c c c}
\toprule
\midrule
index &  $R_\mathrm{s}$ [{\AA}] &  $\lambda$ & $\zeta$ & absolute SHAP value\\
\midrule
1217 & 4.0 & 1 & 16 & 0.0152 \\
1216 & 4.0 & 1 & 8 & 0.0088 \\
1232 & 5.0 & 1 & 32 & 0.0045 \\
1245 & 6.0 & 1 & 16 & 0.0020 \\
1192 & 2.0 & -1 & 1 & 0.0019 \\
\midrule
\bottomrule
\end{tabular}
\label{table:SHAP_Na-Cl-O}
\end{table}

\begin{figure}[t]
    \centering
    \includegraphics[width=0.8\linewidth]{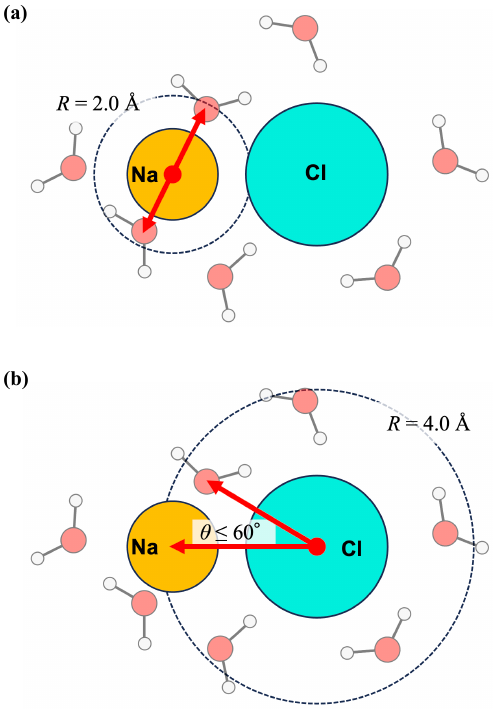}
    \caption{Schematic illustration of  water O atoms 
 within 2 {\AA} of Na (a) and water O atoms within the 4 {\AA} hydration
 shell of Cl (b), characterized by two ACSF descriptors, $G^5_{58}$ and
 $G^5_{1217}$, respectively.
}
     \label{fig:water_bridge}
\end{figure}

\begin{figure}[t]
    \centering
    \includegraphics[width=\linewidth]{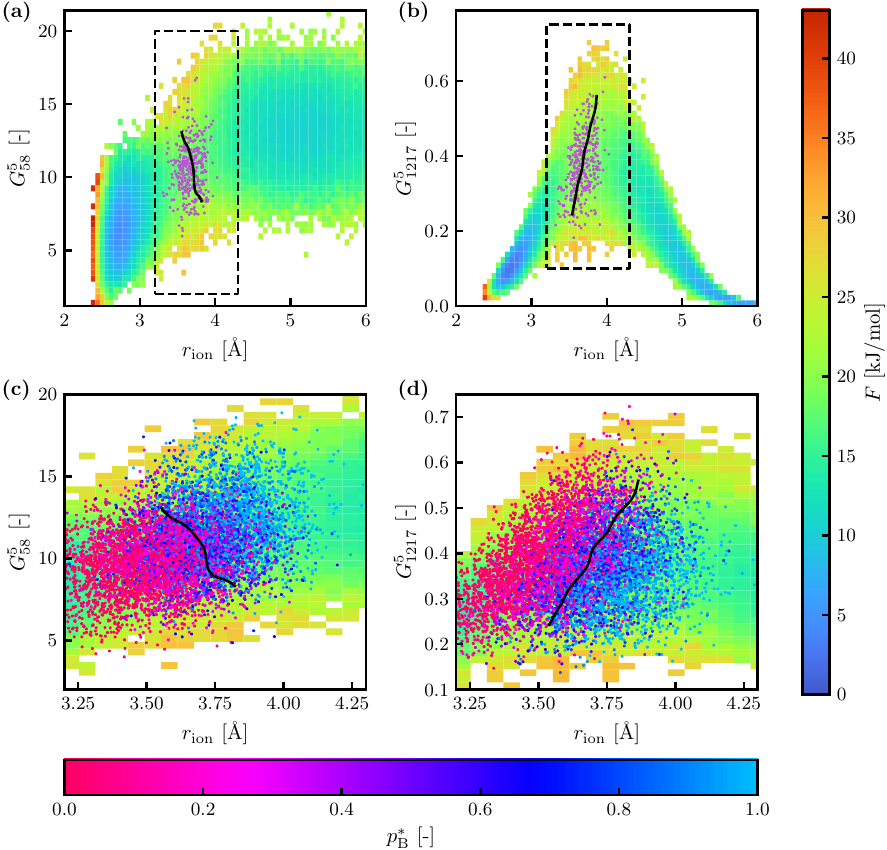}
    \caption{Two-dimensional PMF using the interionic
 distance $r_\mathrm{ion}$ and $G^5_{58}$ (a) and $G^5_{1217}$ (b).
The point with the minimum energy is set to 0 kJ/mol, and values are
 color-coded according to the color bar on the right.
Purple dots indicate configurations with committer values in the range $0.45 < p_\mathrm{B}^* < 0.55$.
The distributions of committor $p_\mathrm{B}^*$ within the rectangular boxes in (a)
 and (b) are described in (c) and (d), respectively.
Committer values are colored according to the bottom color bar.
In each panel, 
the black contour line was obtained as follows: 
within the region in (c) or (d), corresponding the rectangular box
 in (a) or (b),
the ranges of the $x$- and $y$-axis values were divided into 200
 grid points each (forming a 200 $\times$ 200 grid), 
and the average committor value $p_\mathrm{B}^*$ within each grid cell
 was computed. 
After applying cubic interpolation for smoothing, the contour line
 corresponding to $p_\mathrm{B}^*=0.5$ was drawn.
}
     \label{fig:rion_G5}
\end{figure}

\begin{figure}[t]
    \centering
    \includegraphics[width=\linewidth]{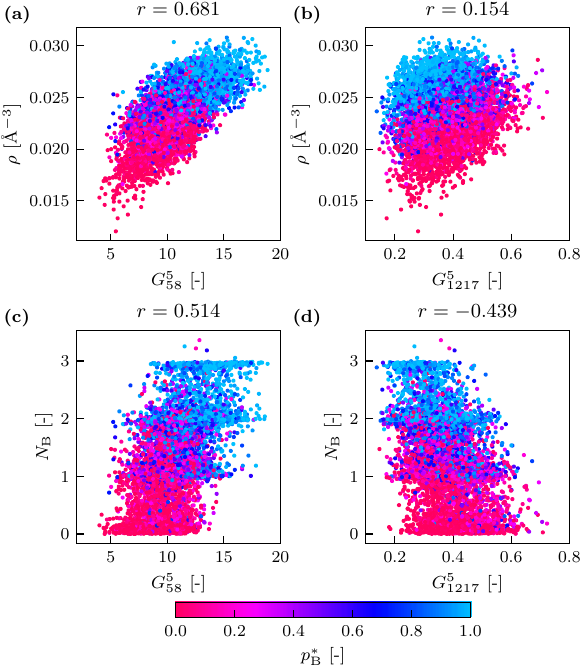}
    \caption{Distribution of committor $p_\mathrm{B}^*$ dataset in the
 two-dimensional surface 
using the
combinations $(G^5_{58}, \rho)$ (a), $(G^5_{1217}, \rho)$ (b),
$(G^5_{58}, N_\mathrm{B})$ (c), and $(G^5_{1217}, N_\mathrm{B})$ (d).
Committer values are colored according to the bottom color bar.
The $r$-value represents the correlation coefficient for each panel.}
     \label{fig:G5_rho_NB}
\end{figure}

\subsection{Feature contribution of CV by SHAP}

The feature contribution of each input CV to the RC predicted by the neural
network was evaluated using SHAP.
The average of 100 SHAP values randomly sampled from the
test dataset of 3040 configurations was calculated and the results are plotted
in Fig.~\ref{fig:SHAP}.

Figure~\ref{fig:SHAP}(a) shows the summed SHAP values for each atomic combination in
the $G^2$ and $G^5$ descriptors.
Overall, the contribution of $G^2$ is minimal, indicating that the
information encoded in $G^2$ is effectively subsumed by the $G^5$
descriptors.
Notably, within $G^5$, ($Z_1$-$i$-$Z_2$)=(O–Na–O) descriptors exhibited the largest
contribution, followed by (Na–Cl–O), (Cl–Na–O), and (H-Cl-H).
This result indicates that the coordination of water O atoms around 
Na ion, as well as configurations involving one ion and a water O
atom relative to the other ion, play a central role.
Note that the environment of water O atoms surrounding both ions is
common to the (Na–Cl–O) and (Cl–Na–O) descriptors.
In contrast, the contribution of water H atoms around Cl ion is
minor.
This relatively minor contribution of solvation 
around Cl ion is consistent with the small change in the
coordination number around Cl ion between the associated and dissociated 
states, as reported by Geissler \textit{et
al.}~\cite{geissler1999Kinetic}
Other descriptors, such as (Cl-Na-H) and (Na-Cl-H), which characterize water H atom
around both ions, and (O-Na-H) and (O-Cl-H), which describe
environments where O and H atoms coordinate around a single ion
are found to be less important.
These results suggest that the arrangement of H atoms around the ions is
less significant than that of O atoms.
Among the ten atom combinations, the (H-Na-H) and (O-Cl-O)
descriptors, representing the coordination of water H and O atoms around Na
and Cl ions, relatively, show the smallest contributions, consistent
with the underlying electrostatic interactions.

Table~\ref{table:SHAP_O-Na-O} lists the top five contributors
the within $G^5$ (O–Na–O) descriptors 
along with their
corresponding absolute SHAP values.
The $G^5$ descriptor at index 58 (denoted as $G^5_{58}$), corresponding to (O–Na–O) with the
highest contribution, has the distance and angular sensitivities of
$R_\mathrm{s}=2.0$ {\AA}, $\lambda=-1$, and $\zeta=1$. 
This indicates that water O atoms within a spherical shell of radius 
$R_\mathrm{s}=2.0$ {\AA} around Na make a significant contribution.
This $G^5_{58}$ is largely consistent with the 
$G^5$ (denoted as $x_7$ in Ref.~\onlinecite{jung2023Machineguided}) with
$R_\mathrm{s}=1.0$ {\AA}, $\zeta=2$ and $\lambda=-1$ identified by Jung \textit{et al.}
Note that 
$R_\mathrm{s}=2.0$ {\AA}, which is slightly larger than that of $x_7$,
approximately corresponds to the Lennard-Jones
diameter of Na ion.
In contrast, the low
angular resolution ($\zeta=1$) suggests that 
angular dependence is not particularly important for characterizing the
distribution of O atoms around Na ion.
Typical structures are schematically illustrated in
Fig.~\ref{fig:water_bridge}(a).
The other descriptors listed in Table~\ref{table:SHAP_O-Na-O} exhibit
similar properties, while their parameter values differ slightly.

In addition to $G^5$ (O–Na–O) descriptors, 
$G^5$ (Na-Cl-O) descriptors also make significant contributions to the RC.
As listed in Table~\ref{table:SHAP_Na-Cl-O}, 
index 1217 (denoted as
$G^5_{1217}$) contributes the most
significantly, characterized by $R_\mathrm{s}=4.0$ {\AA}, $\lambda=1$, and $\zeta=16$.
At this high angular sensitivity ($\zeta=16$), the angular component
$2^{1-\zeta}(1+\cos\theta_{ijk})^{\zeta}$ 
decreases from unity and approaches zero near $\theta_{ijk}\approx
60^\circ$, 
indicating that the dominant Na–Cl–O bond angles lie within this range.
This implies that water O atoms located in the overlapping region of the 
hydration shells, corresponding to $R_\mathrm{s}=4$ {\AA} separation
between Na and Cl ions, slightly larger than the saddle point
distance $r_\mathrm{ion}=3.6$ {\AA} of the PMF, 
are key contributors [see the schematic illustration in Fig.~\ref{fig:water_bridge}(b)].
The other descriptors listed in Table~\ref{table:SHAP_Na-Cl-O} also
imply similar structural features, except for index 1192.

It is important to characterize
the $p_\mathrm{B}^*$ dataset using the interionic distance,
$r_\mathrm{ion}$, in combination with either of 
the two ACSF descriptors, $G^5_{58}$ or $G^5_{1217}$, as shown in
Fig.~\ref{fig:rion_G5}.
Figures~\ref{fig:rion_G5}(a) and \ref{fig:rion_G5}(b)
display the two-dimensional PMFs constructed using ($r_\mathrm{ion}$, $G^5_{58}$)
and ($r_\mathrm{ion}$, $G^5_{1217}$), respectively.
The PMF was calculated using the mbar\_analysis
code~\cite{matsunaga2022Use} in the GENESIS
software,~\cite{jung2015GENESIS, kobayashi2017GENESIS, jung2024GENESIS}
which implements
the Multistate Bennett acceptance ratio
method.~\cite{shirts2008Statistically}
To examine the behavior near the TS, the distributions of
$p_\mathrm{B}^*$ within the rectangular regions shown in Fig.~\ref{fig:rion_G5}(a) and
\ref{fig:rion_G5}(b) are 
presented in Figs.~\ref{fig:rion_G5}(c) and \ref{fig:rion_G5}(d), respectively.
These results demonstrate that 
the committor evolves 
continuously from the associated state
($p_\mathrm{B}^*\approx 0$) to 
the dissociated state ($p_\mathrm{B}^*\approx 1$), with the TS
($p_\mathrm{B}^*= 0.5$) forming clear separatrix lines, which are
consistent with the corresponding PMF profiles.

Specifically, Figs.~\ref{fig:rion_G5}(a) and \ref{fig:rion_G5}(c) show that 
an increase in $G^5_{58}$ accompanied by an increase in
$r_\mathrm{ion}$ leads to the
crossing of the separatrix line, 
indicating 
that enhanced coordination of water O atoms around Na ion plays a crucial
role for the ion pair dissociation.
This observation is consistent with the finding by Geissler \textit{et al.} that the
coordination number around the Na ion increases during the transition
from the associated to the dissociated state.~\cite{geissler1999Kinetic}
Furthermore, Figs.~\ref{fig:rion_G5}(b) and \ref{fig:rion_G5}(d)
demonstrate that, 
starting from the associated state ($p_\mathrm{B}\approx 0$),
$G^5_{1217}$ initially increases together with 
$r_\mathrm{ion}$ as the system approaches the TS ($p_\mathrm{B}=0.5$). 
Upon crossing the separatrix line, however, $G^5_{1217}$ decreases as 
the system proceeds toward the dissociation state ($p_\mathrm{B}\approx 1$).
This behavior
indicates 
that ion pair dissociation requires a reduction in the density of water
O atoms that 
belong simultaneously to the hydration shells of both Na and Cl ions.
Therefore, the two ACSF descriptors, $G^5_{58}$ and $G^5_{1217}$, identified
through the neural network and SHAP analysis, characterize 
the underlying mechanism of the dissociation process of NaCl ion pair
in water.
In particular, the separatrix lines shown in Fig.~\ref{fig:rion_G5}
imply that 
barrier crossing at the TS requires water molecules
coordinating Na ion to migrate toward Cl ion, as pointed out by
Geissler \textit{et al.}~\cite{geissler1999Kinetic}

\subsection{Comparative analysis with CVs representing water bridging structure}

Finally, it is also important to examine the correlations of $G^5_{58}$ and
$G^5_{1217}$ with CVs previously identified using the Inertial
Likelihood Maximization method by M\"{uller} \textit{et al.}~\cite{mullen2014Transmission}
As described in the Introduction, 
the interionic water density $\rho$ and 
and the number of water molecules that simultaneously coordinate both
ions, $N_\mathrm{B}$
were found to contribute most significantly to the RC.
Specifically, the interionic water density $\rho$ is defined by 
\begin{equation}
\rho=\left(\frac{1} {2\pi \sigma^2}\right)^{3/2}
\sum_{w}
\exp
\left(
-\frac{|\bm{r}_w-\bm{r}_\mathrm{mid}|^2}{2\sigma^2}
\right),
\end{equation}
where $\bm{r}_w$ and $\bm{r}_\mathrm{mid}$ represent the
center-of-mass position of the $w$-th water molecule and the midpoint
between Na and Cl ions, respectively.
The parameter $\sigma$ controls the number of water molecules within the volume
$(2\pi \sigma^{3/2})$, which is chosen as $r_\mathrm{ion}/2$.
The number of water molecules that simultaneously coordinate both ions,
$N_\mathrm{B}$, 
is calculated based on the following two steps:
The ion coordination function is defined as
\begin{equation}
f_{s-w} = \frac{1-\tanh[a (R_\mathrm{s-w}-b)]}{2},
\end{equation}
where the subscript $s$ represents an ion species, \textit{i.e.}, Na or
Cl.
The distance $R_\mathrm{s-w}$ corresponds to the $w$-th water O atom when
calculating Na coordination, and to the $w$-th water H atom when calculating
Cl coordination.
The parameters $a$ and $b$ are set to 3 {\AA}$^{-1}$ and $3.2$ {\AA}, respectively.
The expression of $N_\mathrm{B}$ is given by 
\begin{equation}
N_\mathrm{B}  = \sum_{w} \min (f_{\mathrm{Na}-w},
 f_{\mathrm{Cl}-w}), 
\end{equation}
which is thereby regarded as the number of bridging water molecules.

Figure~\ref{fig:G5_rho_NB} shows the distribution of the
committor $p_\mathrm{B}^*$ dataset on the two-dimensional surfaces using the
combinations $(G^5_{58}, \rho)$ (a), $(G^5_{1217}, \rho)$ (b),
$(G^5_{58}, N_\mathrm{B})$ (c), and $(G^5_{1217}, N_\mathrm{B})$ (d).
For each combination of variables, the corresponding correlation
coefficient $r$ was calculated and is presented in Fig.~\ref{fig:G5_rho_NB}.

The interionic water density $\rho$ shows a positive correlation with
$G^5_{58}$ and a weak negative correlation with $G^5_{1217}$.
Note that
the low sensitivity of
$G^5_{58}$ to angular components coincides with the fact that $\rho$ does not
account for the orientation of water molecules.
The two-dimensional PMF using $(r_\mathrm{ion}, \rho)$, reported in
Ref.~\onlinecite{mullen2014Transmission}, indicates that 
an increase in $\rho$ promotes dissociation
from the associated state.
This behavior is also consistent with 
the separatrix line described in Fig.~\ref{fig:rion_G5}(a), which indicates
the increase in $G^5_{58}$ facilitates the ion dissociation process.

In contrast, 
the number of bridging water molecules $N_\mathrm{B}$ exhibits a
positive correlation with $G^5_{58}$, while its correlation with
$G^5_{1217}$ is slightly stronger than that of $\rho$.
Note that a water molecule must adopt an appropriate orientation to contribute
to both $N_\mathrm{B}$ and $G^5_{1217}$.
The two-dimensional PMF using $(r_\mathrm{ion}, N_\mathrm{B})$, reported in
Ref.~\onlinecite{mullen2014Transmission}, suggests the existence of
multiple transition pathways:
each pathway corresponds to $N_\mathrm{B}$=0, 1, or 2, and
as $N_\mathrm{B}$ increases, the $r_\mathrm{ion}$ occupying the TSs
decrease to approximately 4.2, 3.9, and 3.6, {\AA}, respectively.
However, as shown in Fig.~\ref{fig:rion_G5}, such multiple pathways
cannot be directly identified from $G^5_{58}$ and $G^5_{1217}$.
This is due to the discrete nature of $N_\mathrm{B}$, which is effectively
coarse-grained when represented by ACSFs.
Nevertheless, 
the correlations shown in Fig.~\ref{fig:G5_rho_NB} indicates that both
ACSFs, $G^5_{58}$ and $G^5_{1217}$,
effectively represent the hand-crafted CV that describe the water bridging structure
between ions.

\section{Conclusions}

In this study, 
we identified the RC for the NaCl ion dissociation–association process
using deep learning with the committor $p_B^*$ as the training target. 
The committor $p_B^*$ values were obtained by initial configurations obtained
via umbrella sampling constrained along the interionic distance
$r_{\mathrm{ion}}$, followed by 1 ps MD trajectories. 
To describe the solvent environment around the ions, we introduced ACSFs,
$G^2$ and $G^5$, as CVs and used as input variables to the neural network. 
The model was trained by minimizing the cross entropy so that the
relationship between $p_B^*$ and the RC $q$ followed the sigmoidal
function, $p_\mathrm{B}(q)=(1+\tanh(q))/2$.

To interpret the resulting RC $q$, we applied SHAP, an XAI technique.
This analysis revealed that the ACSF descriptors, $G^5_{58}$ and
$G^5_{1217}$, characterizing the water O atom environment around Na ion and
the water O atoms within the hydration shell of Na and Cl ions, respectively
make dominant contributions to the RC.
Notably, the SHAP analysis highlights $G^5_{58}$
as a dominant descriptor, in agreement with the governing relationship
identified by the symbolic regression analysis.~\cite{jung2023Machineguided}
This consistency supports the physical relevance of the learned
functional form for the RC.
Importantly, the functional form described by the symbolic regression
analysis 
is consistent with the input features identified by SHAP, providing
mutual validation between the two analyses.
Additionally, the SHAP analysis clarified that $G^5_{1217}$, 
characterizing 
the O atoms within the hydration shells of Na and Cl, also 
makes a dominant contribution, which has not been emphasized in prior studies.
We further demonstrated that these two ACSF descriptors effectively capture
the water bridging structure essential to ion dissociation, as evidenced
by their correlations with 
conventionally defined CVs such as
the interionic water density $\rho$ and 
number of bridging water molecules $N_\mathrm{B}$.~\cite{mullen2014Transmission}

Although, as outlined in the Introduction, Ballard and Dellago showed
that the inclusion of the third solvation shell is required to
accurately reproduce a
committor distribution peaked around $p_\mathrm{B}^*$,~\cite{ballard2012Mechanism}  
$G^5_{58}$ and $G^5_{1217}$
contains information only for water molecules
in the first solvation shell.
Our candidate CVs included ACSFs with distance sensitivity
up to $R_\mathrm{s}=9.0$ {\AA}, approximately corresponding to the third
hydration shell, but these descriptors were not identified by SHAP as
major contributors.
This suggests that rearrangements of water molecules in the first
hydration shell effectively effectively incorporates the influence of
the second and third coordination shells.

The RC obtained through our integrated approach, which combines
committor based deep learning, ACSF descriptors, and XAI, provides a
refined molecular picture of solvent effects. 
Moreover, this methodology offers a general strategy applicable to a wide
range of condensed phase reactions in which solvent reorganization plays
a crucial role. 
It can be extended to processes such as ligand
binding, nucleation, and biomolecular conformational changes, where the
solvent environment significantly influence the
TS.

\section*{Supplementary material}

The supplementary material includes figures showing the epoch
dependent loss
functions for training and test datasets (Fig.~S1) as well as the distribution of
committor $p_\mathrm{B}^*$ within the range $|q|<\alpha$, where $\alpha$
is used 
as the threshold (Fig.~S2).
It also provides 
a CSV file containing 
the 
definitions of input variables and those ACSF parameters, 
reference atom $i$, $Z_1$, $Z_2$, 
$R_\mathrm{s}$,
$\lambda$, and $\zeta$ (Tables~S1).

\begin{acknowledgments}
The authors acknowledge Professors Yasutaka Kitagawa,  Ryohei Kishi, and
 Kento Kasahara for
 valuable comments.
This work was supported by 
JSPS KAKENHI Grant-in-Aid 
Grant Nos.~\mbox{JP23K23858}, \mbox{JP25H02299}, 
\mbox{JP23K23303}, \mbox{JP23KK0254}, \mbox{JP24K21756},
\mbox{JP25H02464}, \mbox{JP25K02235}, 
\mbox{JP25K00968}, \mbox{JP24H01719}, 
 \mbox{JP22K03550}, and \mbox{JP23H02622}.
We acknowledge support from
the Fugaku Supercomputing Project (Nos.~\mbox{JPMXP1020230325} and \mbox{JPMXP1020230327}) and 
the Data-Driven Material Research Project (No.~\mbox{JPMXP1122714694})
from the
Ministry of Education, Culture, Sports, Science, and Technology
and by
Maruho Collaborative Project for Theoretical Pharmaceutics.
The numerical calculations were performed at Research Center of
Computational Science, Okazaki Research Facilities, National Institutes
of Natural Sciences (Projects: 25-IMS-C052, 25-IMS-C105, 25-IMS-C227).
\end{acknowledgments}

\section*{AUTHOR DECLARATIONS}

\section*{Conflict of Interest}
The authors have no conflicts to disclose.

\section*{Data availability statement}

The data that support the findings of this study are available from the
corresponding author upon reasonable request.

%

%

\end{document}


\setcounter{equation}{0}
\setcounter{figure}{0}
\setcounter{table}{0}
\setcounter{page}{1}

\noindent{\bf\Large Supplementary Material}
\vspace{5mm}
\begin{center}
\textbf{\large  Deep learning of committor for ion dissociation and interpretable
analysis of solvent effects using atom-centered symmetry functions
}
\\

\vspace{5mm}

{Kenji Okada,$^1$ Kazushi Okada,$^1$  Kei-ichi Okazaki,$^{2, 3}$
 Toshifumi Mori,$^{4, 5}$ Kang Kim,$^1$
and Nobuyuki Matubayasi$^1$}
\\

\vspace{5mm}

\noindent
\textit{
$^{1)}$Division of Chemical Engineering, Graduate School of Engineering
 Science, The University of Osaka, Osaka 560-8531, Japan}\\
\textit{
$^{2)}$Research Center for Computational Science, Institute for Molecular Science, Okazaki, Aichi 444-8585, Japan}\\
\textit{
$^{3)}$Graduate Institute for Advanced Studies, SOKENDAI, Okazaki, Aichi 444-8585, Japan}\\
\textit{
$^{4)}$Institute for Materials Chemistry and Engineering, Kyushu University, Kasuga, Fukuoka 816-8580, Japan}\\
\textit{
$^{5)}$Interdisciplinary Graduate School of Engineering Sciences, Kyushu University, Kasuga, Fukuoka 816-8580, Japan}

\end{center}

\begin{figure}[H]
    \centering
\includegraphics[width=0.3\linewidth]{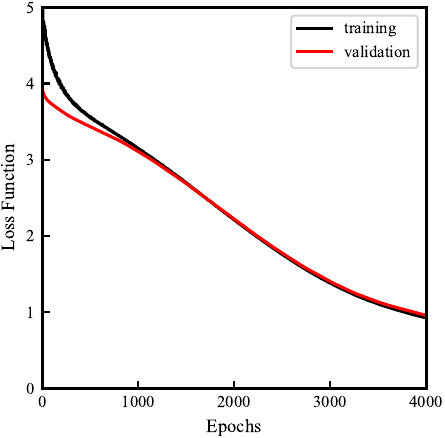}
    \caption{Epoch dependent loss functions of training and test datasets.
}
\label{fig:loss}
\end{figure}

\begin{figure}[H]
    \centering
\includegraphics[width=0.6\linewidth]{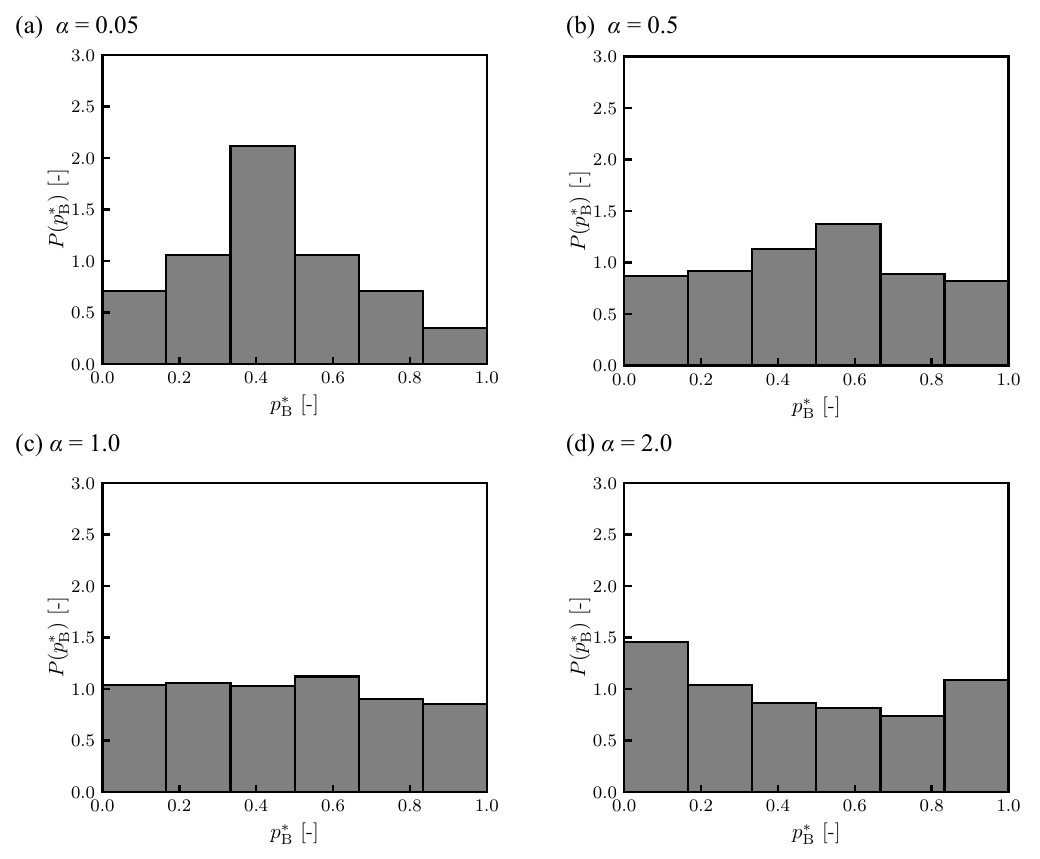}
    \caption{Probability density of committor $p_\mathrm{B}^*$ within
 the range $|q|<\alpha$ with the threshold values, $\alpha=0.05$ (a),
 0.5 (b),
 1.0 (c), and 2.0 (d).}
\label{fig:committor_TS}
\end{figure}